\documentclass[runningheads]{llncs}
\usepackage[utf8]{inputenc}
\usepackage{complexity}
\usepackage{amsmath, amssymb}
\usepackage{macros}
\usepackage{tikz,pgfplots}    
\usetikzlibrary{automata,petri,positioning,shapes,calc,
	decorations.pathmorphing,patterns,arrows,fit,backgrounds,
	shapes.multipart,fadings,shapes.geometric}
\usepackage{thm-restate}
\usepackage{bm}

\usepackage{xcolor}
\definecolor{lightergray}{gray}{0.8}
\definecolor{darkgreen}{rgb}{0.0, 0.6, 0.0}

\usepackage[hidelinks, colorlinks, citecolor=darkgreen]{hyperref}
\usepackage{cleveref}
\usepackage{caption}
\usepackage{subcaption}
\usepackage{colortbl}
\usepackage{enumitem}
\usepackage{stmaryrd}
\usepackage{mathtools}\usepackage[ruled, noend]{algorithm2e} 
\SetKwRepeat{Do}{do}{while}            
\usepackage{thm-restate}

\DeclareMathAlphabet{\pazocal}{OMS}{zplm}{m}{n}

\newcommand{\tr}{\mapsto}

\newcommand{\Stage}{\pazocal{S}}

\newcommand{\WSSS}{\textsc{WSSS}}

\colorlet{colEmph}{black}
\colorlet{colBad}{magenta!60!red}
\colorlet{colGood}{cyan!60!blue}
\colorlet{colNeutral}{gray!70!black}

\newcommand{\AY}{\textcolor{colGood}{\textbf{B}}}
\newcommand{\PY}{\textcolor{colGood}{\textbf{b}}}
\newcommand{\AN}{\textcolor{colBad}{\textbf{R}}}
\newcommand{\PN}{\textcolor{colBad}{\textbf{r}}}

\newcommand{\AYc}{\textcolor{colGood}{\underline{\textbf{B}}}}
\newcommand{\PYc}{\textcolor{colGood}{\underline{\textbf{b}}}}
\newcommand{\ANc}{\textcolor{colBad}{\underline{\textbf{R}}}}
\newcommand{\PNc}{\textcolor{colBad}{\underline{\textbf{r}}}}

\newcommand{\BB}{\mathit{BB}} 
\newcommand{\BBL}{\mathit{BB}_L} 
\newcommand{\STC}{\mathit{STATE}} 

\newcommand{\q}[1]{\mathtt{#1}}                     

\begin{document}
\title{Population Protocols: Beyond Runtime Analysis\thanks{The work surveyed in this note was supported by the  European Research Council (ERC) under the European Union's Horizon 2020 research and innovation program under
grant agreement No~787367 ``Parameterized Verification and Synthesis'' (PaVeS).}}
%
%
\author{Javier Esparza\inst{1}\orcidID{0000-0001-9862-4919}}
\authorrunning{J. Esparza}
%
\institute{Technical University of Munich,  Germany \\
\email{esparza@in.tum.de}
}
\maketitle              

\begin{abstract}
We survey our recent work on the verification of population protocols and their state complexity.
\end{abstract}

\section{Introduction}
Population protocols are a model of computation in which an arbitrary number of
indistinguishable finite-state agents interact in pairs to collectively decide if their initial global configuration satisfies
a given property. Population protocols were introduced by Angluin \etal\ in \cite{AADFP04,AADFP06} to study the theoretical properties of networks of mobile sensors with very limited computational resources, but they are also very strongly related to chemical reaction networks \cite{SCWB08}, a discrete model of chemistry in which agents are molecules that change their states due to collisions.

Population protocols decide a property by \emph{stable consensus}. Each state of an agent is assigned a binary output (yes/no). At each step, a pair of agents is selected uniformly at random and allowed to interact.  
In a correct protocol, all agents eventually reach with probability 1 the set of states whose output correctly answers the question ``did our initial configuration satisfy the property?'' and stay in these states forever.  

The \emph{parallel runtime} of a protocol is defined as the expected number of interactions until a stable consensus is reached (i.e.\ until the property is decided), divided by the number of agents. In recent years, much research on population protocols has focused on the runtime of population protocols, and several landmark results have been obtained. In particular, recent results have studied protocols for majority and leader election in which the number of states grows with the number of agents, and shown that poly-logarithmic time is achievable by protocols without leaders, even for very slow growth functions, see e.g.~\cite{AAEGR17,AlistarhAG18,GasieniecS21}. Many of these results have been described in excellent surveys \cite{AlistarhG18,ElsasserR18}.

My work on population protocols, carried out with many of my PhD students and postdocs, and in collaboration with other colleagues, has focused on other aspects than runtime analysis, and this is the reason for the title of this note. I first learned about population protocols at a talk by Paul Spirakis in ICALP 2008. As a researcher working on
the theory of Petri nets, I noticed the connection of population protocols to Petri nets, and as a researcher working on automatic verification, I asked myself if the correctness problem---given a population protocol and a property, does the protocol decide the property?---was decidable. The problem consisted of checking whether an infinite collection of finite-state Markov chains, one for each initial configuration of the protocol, satisfies a rather sophisticated liveness property with probability~1. This makes it very hard: liveness properties are harder to verify than safety properties, probabilistic systems are harder than non-probabilistic ones, and parameterized problems, i.e., problems involving families of systems with an arbitrarily large number of agents, are \emph{much} harder to verify than systems with a fixed number of agents. After looking at the problem for some time I could not find an answer, but I kept it at the back of my mind, and in 2015 (seven years later!) I suggested to my colleagues Pierre Ganty, J\'er\^ome Leroux, and Rupak Majumdar to examine it again. This time, thanks to new progress by Leroux on the theory of Petri nets, we proved that the correctness problem is decidable, although as hard as the reachability problem for Petri nets \cite{EGLM17}. This was the starting point of a research program devoted to the theory and practice of verifying population protocols, which reached an important milestone in 2020 with the release of \textsc{Peregrine} 2.0, a verifier based on new theoretical results \cite{BEHKM20,EsparzaHJM20}. The first part of this note surveys this research, adding all the work carried out since 2017 to a brief previous survey \cite{Esparza17}. 

In 2018, Michael Blondin, Stefan Jaax and myself observed that a well-known result of the theory of Petri nets had a surprising application to the theory of population protocols: It
showed that an infinite family of predicates of the form $x \geq k$ for certain values of $k$ could be decided by extremely succinct protocols with only $O(\log \log k)$ states. This sparked our interest in the question of how many states are needed to decide a given predicate, or, by analogy to automata theory, the \emph{state complexity} of the predicate. The question is relevant. For example, the fast protocol for majority implicitly described in \cite{AngluinAE08a} has tens of thousands of states. This is an obstacle to implementations of protocols in chemistry, where the number of states corresponds to the number of chemical species participating in the reactions. The number of states is also important because it plays the role of memory in sequential computational models. Indeed, the total memory available to a population protocol is the logarithm of the number of states multiplied by the number of agents. 

To the best of our knowledge, we are the first group to study the state complexity of predicates. While we do not have a complete characterization yet, we have already proved several results.  In 2018 the only bounds on state complexity were  the ones derived from the synthesis procedures of \cite{AAE06,AngluinAE08a}. The input to these procedures is a boolean combination of atomic predicates of the form
$\sum_{i=1}^k a_i x_i \sim b$, where $a_i$ and $b$ are integers, and
$\sim$ is either $<$ or $\equiv_m$, the latter denoting congruence modulo $m$. (It is known that every predicate decidable by population protocols is of this form.) The bounds of \cite{AAE06,AngluinAE08a} are exponential in both the number of atomic predicates, and in their size, with numbers written in binary. Since 2018 we have shown that every predicate has a protocol with a polynomial number of states both in the number and the size of the atomic predicates \cite{BEJ18,BEGHJ20}, and that, as mentioned above, some predicates have much smaller protocols \cite{BEJ18}. Very recently, we have also obtained lower bounds for the state complexity \cite{CzernerE21}. The second part of the note surveys this work.

The note is structured as follows. Section \ref{sec:term} introduces terminology, and Sections \ref{sec:verification} and \ref{sec:succinct} survey our work on verification and state complexity, respectively.

\section{Some terminology}
\label{sec:term}

We assume that the reader is familiar with the basics of the population protocol model; here we just fix some terminology. 

\paragraph{Population protocols.} 
A population protocol has a set of states and transitions, with a distinguished set of initial states. Every state also has 
an associated output, $1$ or $0$. Transitions model interactions between two agents. They have the form $q_1, q_2 \mapsto q_1', q_2'$, meaning that two agents in states $q_1$ and $q_2$ interact and move to states $q_1'$ and $q_2'$. We assume that there is exactly one transition for each pair of states, but transitions can also be silent, meaning that the states of the agents do not change. A configuration of a protocol is a mapping assigning to each state a number of agents. Initial configurations put arbitrarily many agents in the initial states, and $0$ agents elsewhere. A protocol starts at some initial configuration, and executes steps by repeatedly picking two agents uniformly at random and applying the corresponding transition. A run is an infinite sequence of configurations obtained by executing infinitely many steps.

A configuration has consensus $1$ resp. $0$ if all its agents occupy states with output $1$ resp. $0$. We also say that the configuration is a $1$-consensus, resp. a $0$-consensus, or just a consensus when we are not specific.  A configuration is a \emph{stable} $1$-consensus if every configuration reachable from it, including itself, is a $1$-consensus.
A protocol is \emph{well specified} if for every initial configuration $C$ there is $b \in \{0, 1\}$ such that runs starting at $C$ eventually reach a stable $b$-consensus with probability 1 (abbreviated as w.p.1 in this note); in this case, we say that the protocol \emph{outputs $b$} for the initial configuration $C$. 
A well specified protocol with initial states $q_1, \ldots, q_k$  \emph{decides} the predicate $\varphi(x_1, \ldots, x_k)$ defined by: $\varphi(n_1, \ldots, n_k)=b$ if{}f the protocol outputs $b$ for the initial configuration that puts $n_1, \ldots, n_k$ agents in $q_1, \ldots, q_k$.

We also consider protocols \emph{with (one or more) leaders}. Intuitively, this is a population protocol with a set of distinguished agents. Formally, a protocol with leaders only differs from a normal population protocol in the definition of the set of initial configurations. The initial configurations of normal protocols put arbitrarily many agents in the initial states, and $0$ agents elsewhere. In a protocol with leaders the initial configurations  also put a fixed number of agents, the same for every initial configuration, in other states.

Population protocols can be seen as special classes of Petri nets or Vector Addition Systems \cite{Reisig85a,murata1989petri}. For the purposes of this note, it suffices to say that, like a protocol, a Petri net has a set of states (called \emph{places} in Petri net jargon) and transitions. However, transitions have the form 
$q_1, \ldots, q_n \mapsto q_1', \ldots, q_m'$ for arbitrary $n, m \geq 0$. So, intuitively, transitions of a Petri net can represent interactions between more than two agents, and they can create or destroy agents. The theory of Petri nets has produced numerous results about the properties of their reachability graphs, i.e., the graphs with the configurations as nodes, and the steps as transitions. Such results can be immediately translated to population protocols. 

\paragraph{Presburger arithmetic, Presburger predicates, and Presburger sets.}\label{sec-presburger-formulas}

A fundamental result of the theory of population protocols is that they decide precisely the Presburger predicates, i.e., the predicates expressible in Presburger arithmetic~\cite{AAER07}.  We briefly recall the definition  of Presburger arithmetic and some results, and refer to \cite{Haase18} for more details.

Atomic formulas of Presburger arithmetic are of the form
$\sum_{i=1}^k a_i x_i \sim b$, where $a_i$ and $b$ are integers, $x_i$ are variables, and
$\sim$ is either $<$ or $\equiv_m$, the latter denoting the congruence modulo $m$ for some $m \ge 2$. We call these atomic formulas \emph{threshold} and \emph{remainder} (or \emph{modulo}) formulas, respectively. The formulas of Presburger arithmetic are the result of closing atomic formulas under Boolean operations and first-order existential or universal quantification. 
For example, $$\varphi(x, y) = \forall z \, \exists u  \colon  3x - u \leq 0 \wedge 2z - y + u \geq  4 \wedge (x+y) \equiv_5 3$$
\noindent is a formula of Presburger arithmetic with free variables $x$ and $y$. 

A formula $\varphi(x_1, \ldots, x_n)$ is interpreted over $\Nat^n$ in the expected way. A set of vectors $S \subseteq \Nat^n$ is a \emph{Presburger set} if there is a Presburger formula $\varphi$ such that a vector belongs to $S$ if{}f it satisfies $\varphi$, and a predicate over  $\Nat^n$ is a \emph{Presburger predicate} if the set of vectors satisfying the predicate is a Presburger set. So Presburger formulas are finite representations of the Presburger sets and predicates. \emph{Semilinear sets} are another representation of the Presburger sets, that is, a set is semilinear if{}f it is Presburger. In this note we do not need any specific properties of the semilinear representation.

Presburger arithmetic has a quantifier-elimination procedure, meaning that every formula can be transformed into an equivalent boolean combination of threshold and remainder predicates. The satisfiability problem for full Presburger arithmetic is decidable, but its complexity is high, it lies between  2-NEXP and 2-EXPSPACE. For quantifier-free formulas the problem is NP-complete, and there exist SMT-solvers efficient in practice. The tools \textsc{Peregrine} 1.0 and 2.0 described later in this note are implemented on top of the Z3 solver~\cite{MouraB08}.

\section{Verification of population protocols}
\label{sec:verification}

The design of population protocols is quite error prone. In our experience, it is hardly ever the case that the first design for a protocol computing a predicate is correct. The problem is accentuated by the lack of a suitable high-level programming language for protocols, which makes their design akin to writing machine code. 

In this context, the limited expressive power of population protocols also has a positive side: the correctness problem is not trivially undecidable, as happens with many other models of computation. In this section we show that the problem is in fact decidable, and survey our work leading from the first decidability result to the current decision procedures and to their practical implementation.

\subsection{Decidability and complexity}
In \cite{EGLM17} we proved that the two central verification problems for population protocols are decidable:
\begin{itemize}
\item Well-specification: Given a population protocol, is it well specified? 
\item Correctness: Given a population protocol and a Presburger predicate (represented as a Presburger formula), does 
the protocol compute the predicate? 
\end{itemize}
\noindent The results were extended to decidability of more general properties in \cite{EGLM16}. The proofs proceed by reduction to the reachability problem for Petri nets, which is decidable \cite{Mayr81,Kosaraju82}. We also showed that well-specification and correctness are recursively equivalent to the reachability problem for Petri nets. More precisely:
\begin{itemize}
\item The reachability problem for Petri nets can be reduced to well-specification or correctness in polynomial time;
\item Given an oracle for the reachability problem for Petri nets, well-specification and correctness can be decided in elementary time, i.e., in time bounded by a finite tower of exponentials. 
\end{itemize}
It has been shown recently that the reachability problem for Petri nets is Ackermann-complete, meaning that its complexity is bounded from below and from above by non-primitive recursive fast growing functions related to the Ackermann function \cite{CzerwinskiLLLM21,lerouxAck,czerwinskiAck}. Therefore, well-specification and correctness  are not primitive recursive either, that is, no algorithm running in time bounded by a primitive recursive function can solve them.


The very high complexity of the correctness problem leads to the question whether the problem could be more tractable for special protocol classes. In \cite{AAER07} Angluin \etal\ not only characterized the expressive power of population protocols, but also of other models with more restricted communication mechanisms. In \cite{EsparzaRW19,EsparzaJRW21} we conducted a complete analysis of the complexity of the correctness problem for the models of \cite{AAER07}. We showed that for models based on immediate and delayed observation the correctness problem is PSPACE-complete and complete at the second level of the polynomial hierarchy, respectively. \emph{Immediate observation} protocols have transitions of the form $q_1, q \mapsto q_2, q$. Intuitively, an agent in state $q_1$ \emph{observes} that the other process is in state $q$, which allows it to immediately move to state $q_2$. Intuitively, in such protocols  if an agent at $q_1$ can execute the transition, then every agent at $q_1$ can take it as well, which greatly simplifies the verification task. In \emph{delayed observation} protocols the observer in $q_1$ may move to $q_2$ at a later point.

\subsection{A first attempt at a verification tool} 
In \cite{BEJM17}, published in 2017, we addressed the problem of developing an algorithm that would be efficient enough to automatically prove correctness for a class of protocols satisfying three conditions:
\begin{itemize}
\item[(a)] No loss of expressive power: the class should compute all Presburger predicates.

\item[(b)] Naturality: the class should contain many of the protocols discussed in the literature.

\item[(c)] Feasible membership problem: deciding if a protocol belongs to the class should have reasonable complexity.
\end{itemize}

In the paper we introduced the class of  \emph{Well-Specified Strongly Silent} protocols (\WSSS). 
Intuitively, a protocol is silent if an execution reaches a terminal configuration with probability 1, where a configuration is terminal if cannot reach any other configuration. (Observe that a protocol correctly deciding a property need not be silent; indeed, the definition of when a protocol decides a property only requires  that an execution reaches a \emph{consensus} w.p.1. Reaching a consensus is a weaker property, because it allows the protocol to keep visiting different configurations, as long as in all of them the agents agree on the same value.) Further, a protocol is strongly silent if, loosely speaking, its transitions are organized in layers such that transitions of higher layers cannot be enabled by executing transitions of lower layers. In particular, if the protocol reaches a configuration of the highest layer that does not enable any transition, then this configuration is terminal.  We showed that {\WSSS} protocols satisfy~(a) and~(b), and proved that the membership problem for the class is in the complexity class \DP{}. Recall that \DP{} is the class of
languages $L$ such that $L = L_1 \cap L_2$ for some languages $L_1 \in \NP$ and $L_2 \in \coNP$; in view of the Ackermannian complexity of the general case, this result shows that {\WSSS} satisfies (c). The proof
that the problem is in \DP{} reduces membership to checking satisfiability resp. unsatisfiability of two quantifier-free formulas of Presburger arithmetic. The procedure was implemented
in \textsc{Peregrine} 1.0  \cite{BEJ18b}, a tool for the verification of population protocols built on top of the constraint solver Z3, and the first tool able to automatically prove well-specification for \emph{all} initial configurations. 

While {\WSSS} protocols decide all Presburger predicates, \textsc{Peregrine} 1.0 had several limitations, which will be subject of the next section. From the most conceptual to the most practical:
\begin{itemize}
\item The verification algorithm of \cite{BEJM17} was dissociated from the decidability results proved in \cite{EGLM17}. To put it bluntly, the theoretical result was not guiding the design of a practical algorithm. 
\item The tool did not produce correctness certificates; if the tool returned ``correct'', the user had to trust the result. 
\item Many protocols designed to be fast, or to use few states, are not in {\WSSS}. Examples include the average-and-conquer protocol of \cite{AGV15} (for fixed values of the parameters), or the compact threshold protocols of \cite{BEGHJ20}. In particular, many protocols that perform a ``random search'', like the second protocol in Example \ref{ex:silent} below, are non-silent. 
\end{itemize}

\begin{example}
\label{ex:silent}
The following two (very slow) protocols decide whether the number of blue agents minus the number of red agents is at least $2^k$ for a given $k \geq 1$. These protocols are also of interest in the next section on state complexity.

\paragraph{First protocol.} Each agent has a bag that can hold up to $2^k$ pebbles and a flag that can be up or down (corresponding to output $1$ and $0$, respectively). 
Initially, each agent has one pebble and its flag is down. When two agents meet they update their bags and flags depending on their colors:
\begin{itemize}
\item Two red agents. No change.
\item  One blue and one red agent. If none of the two bags is empty, then both agents throw one pebble away and lower their flags; we call this interaction a \emph{cancellation}. Else, if the bag of the blue agent is full (that is, if it has $2^k$ pebbles) or if both flags were up before the interaction, then both agents raise their flags. Else, both agents lower their flags. 
\item Two blue agents. One agent gives the other as many pebbles as the other agent's bag can still hold. If this fills the other agent's bag, or if both flags were up before the interaction, then both agents raise their flags; otherwise they lower them. 
\end{itemize}

Let us prove correctness. Let $x$ and $y$ be the numbers of blue and red agents, respectively. W.p.1 cancellations keep occurring until a configuration $C$ is reached in which only blue or only red agents have pebbles (or no agent has pebbles).  If $x - y \geq  2^k$, then no red agents have pebbles at $C$, and in runs from $C$ some blue agent fills its bag and raises its flag w.p.1. The bag remains full forever, and w.p.1 this agent eventually meets all other agents \emph{without any other interaction happening in-between}, after which all flags are up, and stay up forever. If $x~-~y~<~2^k$, then after $C$ no bag is ever full, and so any flag that is lowered stays down forever. Moreover, at $C$ the flag of the blue agent that participated in the last cancellation is down, and this agent brings down the flag of any agent it meets. So eventually all flags stay down w.p.1.

\paragraph{Second protocol.} Again, agents have bags and flags. The following updates ensure that the number of pebbles in each bag is always $0$ or a power of $2$:
\begin{itemize}
\item Two red agents. No change.
\item One blue and one red agent. If both agents have \emph{exactly one pebble}, they throw their pebbles away and lower their flags.  
Else, if the bag of the blue agent is full or both flags were up before the interaction, then both agents raise their flags. Else, both agents lower their flags. 
\item Two blue agents. If both agents hold \emph{the same number of pebbles}, then one of them gives to the other as many pebbles as the other's bag can still hold; if one agent has no pebbles, then it receives from the other half its pebbles; otherwise no pebbles are exchanged. If after this some bag is full, or if both flags were up before the interaction, then both agents raise their flags, else they lower them.  
\end{itemize}

This protocol is also correct. Intuitively, blue agents can distribute pebbles among them into any combination of powers of $2$ (up to $2^k$). For example, if $k=2$  the blue agents can partition $5$ pebbles among them as 1+1+1+1+1 (5 agents get one pebble each);  2+1+1+1; 2+2+1; or 4+1. Randomness ensures that all these partitions are visited infinitely often, and so that cancellations keep occurring until a configuration $C$ is reached  in which only blue or only red agents have pebbles (or no agent has pebbles). 

If $x - y \geq  2^k$, then runs from $C$ eventually execute the following sequence of interactions w.p.1: first, some blue agent fills its bag and raises its flag; this agent then proceeds to meet all red agents, and then all blue agents whose bag is not empty; after that, the agent meets each blue agent with empty bag, sharing its pebbles with it, but only to recover them immediately. After this sequence all flags are up, and remain so forever. If $x - y <  2^k$, then the argument is as for the first protocol.

\medskip

The first protocol is silent. If $x \geq y$, then w.p.1 it eventually reaches and stays in the configuration in which $\lfloor (x-y)  / 2^k \rfloor$ blue agents have $2^k$ pebbles, one blue agent has $(x-y)\!\!\mod 2^k$ pebbles each, all other agents have $0$ pebbles, and all flags are up or down, depending on whether $x -y \geq 2^k $ holds or not. If $x \leq y$,  the protocol reaches and stays in the configuration in which $y-x$ red agents have one pebble each, all other agents have $0$ pebbles, and all flags are down. The second protocol needs exponentially fewer states, but is not silent. Indeed, when  $ x\geq y$ the blue agents keep visiting all partitions of $x-y$ forever. 
\end{example}

\subsection{A new proof methodology: stage graphs}

Finding theoretical and algorithmic answers to the limitations of \textsc{Peregrine} 1.0 took three years. Initially we did not even have a clear picture of these limitations. They emerged when we started to investigate how to automatically compute an upper bound on the expected runtime of a protocol. This work, published in \cite{BEK18}, introduced stage graphs, a notion that, after many rewrites, finally led to the stage graph proof methodology of \cite{BEHKM20}, which we describe now.

Stage graphs reduce correctness to checking that certain assertions are inductive invariants, and that certain ranking functions decrease in appropriate ways. For standard sequential programs these checks are still undecidable problems, but for population protocols they turn out to be decidable. 

\subsubsection{Certificates of correctness} 
The \emph{reachability graph} of a population protocol has all possible configurations as nodes, and an edge from $C$ to $C'$ whenever $C'$ is reachable from $C$ in one step. It is an infinite graph, but every configuration has finitely many descendants. (A consequence of the fact that transitions do not change the number of agents.)  We call graphs with this property \emph{weakly finite} \cite{EsparzaGK12}. An edge of the graph corresponds to executing one transition of the protocol. The probability of executing a transition at configuration $C$ is the fraction of the pairs of agents at $C$ that enable the transition.  Equipped with these notions, let us briefly review how to certify different kinds of properties. 

\paragraph{Certifying safety.} Safety properties can be certified using inductive invariants. An inductive invariant is just a set of configurations closed under the reachability relation, \ie, if a configuration belongs to the set, then so do all its successors. Given a set $I$ of initial configurations and some set $D$ of dangerous configurations, an inductive invariant $\mathit{Int}$ satisfying $I \subseteq \mathit{Int}$ and $\mathit{Int} \cap D = \emptyset$ certifies that $D$ is not reachable from $I$. 

\paragraph{Certifying termination.} Liveness properties, like termination, can be certified by ranking functions assigning to each configuration an element of a set with a well-founded order, like the natural numbers. Termination for \emph{all} runs of the program starting at $I$ is certified by an inductive invariant $\mathit{Int}$ containing $I$, and a \emph{strictly decreasing} ranking function over $\mathit{Int}$, i.e., a ranking function that strictly decreases whenever the protocol takes a step.


\paragraph{Certifying termination w.p.1.} Termination \emph{with probability 1} can also be certified by an inductive invariant $\mathit{Int}$ containing $I$, and a ranking function $f$.  However, the ranking function only needs to be \emph{weakly decreasing}, meaning that for every non-terminal configuration $C \in \mathit{Int}$, some configuration $C'$ reachable from $C$ in one or more steps satisfies $f(C') < f(C)$. Indeed, if such a function exists, then for every non-terminal configuration of $C \in \mathit{Int}$ there is a positive probability of taking a path that stays within $\mathit{Int}$ by inductivity, and decreases the ranking function. Since the reachability graph is weakly finite, this probability is bounded from below by some $\epsilon > 0$ independent of $C$. So runs reach and stay at terminal configurations w.p.1. 


\paragraph{Certifying stable consensus w.p.1.}  In order  to certify that a run starting at a given set $I$ eventually reaches stable consensus $b$ w.p.1, for some given $b \in \{0,1\}$, we need two inductive invariants $\mathit{Int}_1$ , $\mathit{Int}_2$ and a ranking function $f$ satisfying three properties: 
\begin{itemize}
\item $\mathit{Int}_1$ contains $I$ (and so, by inductivity, also all configurations reachable from $I$) and $\mathit{Int}_2$; 
\item $\mathit{Int}_2$ contains only $b$-consensus configurations (and so, by inductivity, any run reaching $\mathit{Int}_2$ reaches stable consensus); and
\item $f$ is weakly decreasing on  $\mathit{Int}_1 \setminus \mathit{Int}_2$, \ie, for every $C \in \mathit{Int}_1 \setminus \mathit{Int}_2$,  some $C'$ reachable from $C$ in one or more steps satisfies $f(C') < f(C)$. 
\end{itemize}
\noindent The same argument as above shows that runs starting at $C \in \mathit{Int}_1 \setminus \mathit{Int}_2$ eventually reach $\mathit{Int}_2$ w.p.1, and, since  $\mathit{Int}_2$ is inductive, get trapped in $\mathit{Int}_2$. Since $\mathit{Int}_2$ only contains configurations with consensus $b$, runs starting at $I$ reach stable consensus $b$ w.p.1. 

This proof technique is \emph{complete}, meaning that if a run starting at $I$ eventually reaches stable consensus $b$ w.p.1, then we can always find a suitable $\mathit{Int}_1$ and $\mathit{Int}_2$, and  $f$. Indeed, it suffices to choose 
$\mathit{Int}_1$ as the set of all configurations reachable from $I$; $\mathit{Int}_2$ as the set of all configurations of $\mathit{Int}_1$ with stable consensus $b$; and $f$ as the function assigning $0$ to the configurations of $\mathit{Int}_2$, and $1$ to the configurations of $\mathit{Int}_1 \setminus \mathit{Int}_2$.

\paragraph{Stage graphs.}  It is useful to split stable consensus proofs into a small steps.  For this one can exhibit a finite, directed, and acyclic graph, whose nodes are pairs $v =(\mathit{Int}, f)$, where $\mathit{Int}$ is an inductive invariant, and $f$ is a ranking function certifying that runs starting at $\mathit{Int}$ eventually get trapped in $\mathit{Int}_1 \cup \ldots \cup \mathit{Int}_n$ w.p.1, where $\mathit{Int}_1, \ldots, \mathit{Int}_n$ are the invariants of the children of $v$.  Further, the invariants of the bottom nodes of the graph only contain consensus configurations. In \cite{BEHKM20} we call these objects \emph{stage graphs}, and their nodes \emph{stages}. A stage graph proves that runs starting at a stage ``travel down the graph w.p.1'' until they reach a bottom stage, and so stable consensus. Intuitively, stages correspond to ``milestones'' towards stable consensus. 

To prove a protocol correct, we produce two stage graphs proving that runs starting at initial configurations that satisfy the predicate eventually reach stable consensus $1$, and the corresponding property for stable consensus $0$. The stage graphs have an initial stage containing the initial configurations satisfying or not satisfying the predicate, respectively. Observe that, since stages are inductive the initial stages also contain every reachable configuration.  Let us examine stage graphs in more detail with the help of a well-known example. 

\begin{example}
\label{ex:majority}
Consider the following majority protocol, whose purpose is to decide if the initial configuration contains more blue agents than red agents. 
Apart from red or blue, agents can also be active of passive, yielding four possible states $Q = \{ \AY, \allowbreak \AN, \allowbreak \PY, \PN \}$
(uppercase for active agents, lowercase for passive ones). The initial states are $\AY$ and $\AN$, and so initially all agents are active.
The protocol has four transitions:
\begin{alignat*}{5}
    t_1 &\colon & \AY , \AN &\tr \PY , \PN &\qquad\qquad
    t_2 &\colon & \AY , \PN &\tr \AY , \PY \\
    t_3 &\colon & \AN , \PY &\tr \AN , \PN &\qquad\qquad
    t_4 &\colon & \PY , \PN &\tr \PY , \PY \\[-17pt]
\end{alignat*}
\noindent
The blue states $\AY, \PY$ have output $1$, and the red states $\AN, \PN$ output $0$.  So in this case, for better visualization, we call the outputs ``blue'' and ``red'', instead of $1$ and $0$. The protocol is correct if it satisfies the following property: for every initial configuration $C$, \ie, every configuration  $C$ satisfying $C(\PY)=C(\PN)=0$, if $C(\AY) < C(\AN)$, eventually all agents stay forever in the red states $\{\AN, \PN\}$ w.p.1, and if $C(\AY) \geq C(\AN)$  eventually all agents stay forever in the blue states $\{\AY, \PY\}$ w.p.1. 
\end{example}

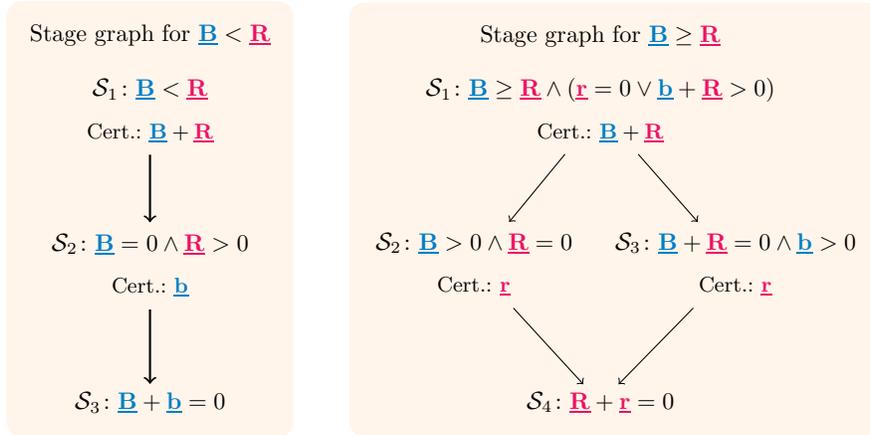
\begin{figure}[h]
\begin{center}
\begin{minipage}{11.7cm}
\pgfdeclarelayer{background}
\pgfdeclarelayer{foreground}
\pgfsetlayers{background,main,foreground}
\begin{tikzpicture}[auto, thick, scale=0.9, transform shape, font=\normalsize]

\tikzset{every node/.style={anchor=north}}


  \node[](n5) {\begin{tabular}{c}$\Stage_1 \colon \AYc < \ANc$\\[0.2cm]
  \footnotesize{Cert.: $\AYc+\ANc$}\end{tabular}} ;
  \node[](n6) [below =1cm of n5] {\begin{tabular}{c}$\Stage_2 \colon \AYc=0 \wedge \ANc > 0$\\[0.2cm]
  \footnotesize{Cert.: $\PYc$}\end{tabular}};
  \node[](n7) [below =1.1cm of n6] {$\Stage_3 \colon \AYc+\PYc=0$};
  
  \path[->]
  (n5) edge node[left]{}(n6) 
  (n6) edge node[left]{} (n7);
  
  \node[](b1) [above =0.2cm of n5] {Stage graph for  $\AYc < \ANc$};
  
  \begin{pgfonlayer}{background}
    \filldraw [line width=4mm,join=round,orange!8]
      (b1.north  -| b1.west)  rectangle (n7.south  -| b1.east);
  \end{pgfonlayer}

  \node[](n0) [right=2.8cm of n5]{\begin{tabular}{c}$\Stage_1 \colon \AYc \geq \ANc \wedge (\PNc = 0 \vee \PYc + \ANc > 0)$\\[0.2cm] \footnotesize{Cert.: $\AYc+\ANc$}\end{tabular}} ;
  \node[](n1) [below left=1cm and -2.5cm of n0, anchor=north east] {\begin{tabular}{c}$\Stage_2 \colon \AYc > 0 \wedge \ANc = 0$\\[0.2cm] \footnotesize{Cert.: $\PNc$}\end{tabular}};
  \node[](n2) [below right=1cm and -2.7cm of n0, anchor=north west] {\begin{tabular}{c}$\Stage_3 \colon  \AYc + \ANc = 0 \wedge \PYc > 0$\\[0.2cm] \footnotesize{Cert.: $\PNc$}\end{tabular}};
  \node[](n3) [below=3.4cm of n0] {$\Stage_4 \colon \ANc+ \PNc = 0$};

  \path[->]
  (n0) edge node[above]{}(n1)
  (n0) edge node[above]{}(n2)  
  (n1) edge node[left]{}(n3)    
  (n2) edge node[left]{} (n3);
  
  \node[](b0) [above =0.2cm of n0] {Stage graph for $\AYc \geq \ANc$};
  \begin{pgfonlayer}{background}
    \filldraw [line width=4mm,join=round,orange!8]
      (b0.north  -| n1.west)  rectangle (n3.south  -| n2.east);
  \end{pgfonlayer}
\end{tikzpicture}
\end{minipage}
\end{center}
\caption{Stage graphs for the protocol of Example \ref{ex:majority}}
\label{fig:stagegraph}
\end{figure}

Figure \ref{fig:stagegraph} shows two stage graphs for the protocol.  Stages are given as constraints over variables
$\AYc, \ANc, \PYc, \PNc$, describing the number of agents in states $\AY, \AN, \PY, \PN$, respectively. For example, the constraint $\AYc < \ANc$ represents the set of all configurations $C$ satisfying $C(\AY) < C(\AN)$. Ranking functions are described as functions of $\AYc, \ANc, \PYc, \PNc$. For example, the function $\AYc+\ANc$ assigns to every configuration $C$ the number $C(\AY)+C(\AN)$. 

The stage graph on the left of  Figure \ref{fig:stagegraph} proves that runs starting at any configuration satisfying $\AYc < \ANc$  reaches stable consensus red with probability 1.  The ``human'' proof goes as follows:  because of transition $t_1$,  from any configuration satisfying $\AYc < \ANc$ the protocol eventually reaches a configuration satisfying $\AYc~=~0$, and then transition $t_3$ eventually changes all remaining blue agents into red agents. The stage graph reflects this proof structure:
\begin{itemize}
\item The initial stage $\Stage_1$ contains exactly the configurations satisfying $\AYc < \ANc$.  The ranking function $f(\AYc, \ANc, \PYc, \PNc) = \AYc + \ANc$ certifies that runs starting at a configuration of $\Stage_1$ eventually get trapped in $\Stage_2$, the set of configurations satisfying $\ANc > 0 \wedge  \AYc=0$, w.p.1. Indeed, consider any configuration  $C~\in~\Stage_1~\setminus~\Stage_2$, i.e., a configuration satisfying $0 < \AYc < \ANc$. Then $C$ enables transition $t_1$.  Letting $C \trans{t_1} C'$ we have  
$$f(C') = \AYc' + \ANc' = (\AYc - 1) + (\ANc -1) < \AYc + \ANc = f(C) \ .$$
\noindent So $f$ is weakly decreasing. 
\item Similarly, the ranking function $g(\AYc, \ANc, \PYc, \PNc) = \PYc$ certifies that runs starting at a configuration of $\Stage_2 \setminus \Stage_3$ eventually get trapped in $\Stage_3$. Since $\Stage_3$ is the set of configurations without blue agents, we are done.  Observe that not every transition decreases $\PYc$; actually, transition $t_4$, which is enabled at some configurations of $\Stage_2 \setminus \Stage_3$, increases it. However, $g$ is weakly decreasing because of transition $t_3$.
\item Observe that $\Stage_1, \Stage_2, \Stage_3$ are inductive invariants. For example, if a configuration satisfies $\AYc < \ANc$, then so does any configuration reached by applying any of the four transitions of the protocol.
\end{itemize}

Let us now consider the stage graph on the right. It proves that runs starting at initial configurations satisfying $\AYc \geq \ANc$ (that is, at the set of configurations satisfying $\AYc \geq \ANc$ and  $\PY=0 = \PN$) reach stable consensus blue w.p.1. The choice of the initial stage $\Stage_1$ is not completely trivial. $\Stage_1$ must satisfy three conditions: (a) contain all configurations satisfying $\AYc \geq \ANc$ and  $\PY=0 =\PN$; (b) contain only configurations with a majority of blue agents or a tie, because only those configurations reach stable consensus blue; (c) be inductive. We cannot choose $\Stage_1 = \AYc \geq \ANc$ because it violates (b); for example, the configuration given by $\AYc=1$, $\ANc=1$, $\PYc=1$, $\PNc=2$ has a majority of red agents. We cannot choose $\Stage_1 = \AYc \geq \ANc \wedge \PY=0 =\PN$ either, because it violates $(c)$. One has to find the right set between these two.

The ``human'' proof uses as milestone the configurations at which there are no agents left in state $\AN$. These configurations can be of three kinds, corresponding to the stages $\Stage_2$, $\Stage_3$, and $\Stage_4$: 
\begin{itemize}
\item if there are no red agents left (stage $\Stage_4$), then the run has already reached stable consensus blue; 
\item if there are agents left in $\AY$ (stage $\Stage_2$), then any agents in state $\PN$ are eventually turned blue by transition $t_2$; 
\item if there are agents left in $\PY$ (stage $\Stage_3$), then any agents in state $\PN$ are eventually turned blue by transition $t_4$.
\end{itemize}

\subsubsection{Decidability of correctness.}

If we are given stage graphs and told they prove that a protocol correctly decides a given property, we can in principle check this statement. We need to check that the initial stages contain the initial configurations satisfying and violating the property, respectively; that all stages are inductive and all ranking functions weakly decreasing; and that the bottom stages only contain configurations with the right consensus. However, since stages are infinite sets, the problem of carrying out such checks might be undecidable. The main theorem of \cite{BEHKM20} proves that, if a protocol is correct, then there exist stage graphs for which the checks reduce to proving satisfiability of formulas of Presburger arithmetic, which is decidable. More precisely, the theorem proves that every correct protocol has \emph{Presburger stage graphs}, \ie, stage graphs satisfying the following properties:

\begin{itemize}
\item Stages are Presburger sets of configurations, i..e, sets expressible in Presburger arithmetic.  
\item Ranking functions are \emph{Presburger functions}. \\
A ranking function $f$ is \emph{Presburger} if there is a formula $\varphi(\mathbf{C}, \mathbf{n})$ of Presburger arithmetic with free variables
$\mathbf{C}$ and $\mathbf{n}$ such that for every configuration $C$ and every number $n$ we have $f(C) = n$ if{}f $\varphi(C, n)$ holds. 
\item Each ranking function $f$ comes equipped with a bound $B$ such that for every configuration $C$ in the domain of $f$,  some configuration $C'$ reachable from $C$ \emph{in at most $B$ steps} satisfies $f(C') < f(C)$.  So, strictly speaking, a Presburger stage graph consists not only of stages and ranking functions, but also of bounds for these functions.
\end{itemize}

Let us now see why checking that a stage graph is Presburger reduces to the satisfiability problem of Presburger arithmetic. 
Checking that the initial stages contain all initial configurations, and that the bottom stages only contain configurations with the right consensus is easy, because the sets of initial configurations and consensuses are Presburger. Let us consider the other two checks.

\paragraph{Checking that a stage is inductive.} Since stages are Presburger sets, given a stage $S$ there is a formula $S(\mathbf{C})$ expressing it. Further, for every transition $t$ it is easy to construct a formula $\mathit{step}_t(\mathbf{C}, \mathbf{C}')$  that holds if{}f $C \trans{t} C'$. For example, for our majority protocol we have
$$\mathit{step}_{t_1}(\mathbf{C}, \mathbf{C}') := \mathbf{C} \geq (1,1,0,0) \wedge  \mathbf{C}'= \mathbf{C} + (-1,-1,1,1)$$
So inductivity  is expressed by the formula 
$$\forall \mathbf{C},\mathbf{C}' \colon \left( S(\mathbf{C}) \wedge \bigvee_{t \in T} \mathit{step}_t(\mathbf{C}, \mathbf{C}')\right) \rightarrow  S(\mathbf{C}')$$
\noindent where $T$ is the set of transitions of the protocol.

\paragraph{Checking that a ranking function is weakly decreasing.}  If the reachability relation of population protocols would be expressible in Presburger arithmetic, i.e., if there were a Presburger formula $\mathit{reach}(\mathbf{C}, \mathbf{C}')$ such that 
$\mathit{reach}(C, C')$ holds if{}f $C \trans{*} C'$, then the weakly-decreasing property for arbitrary ranking functions would be expressible by the formula
$$ \exists \mathbf{C}', \mathbf{n}, \mathbf{n}' \colon \mathit{reach}(\mathbf{C}, \mathbf{C}') \wedge \varphi_f(\mathbf{C},\mathbf{n}) \wedge \varphi_f(\mathbf{C}',\mathbf{n}') \wedge \mathbf{n}' \geq \mathbf{n}$$
However, this is not the case; it is well known that the reachability relation of Petri nets may not be Presburger, and the result easily transfers to population protocols. This is the reason for the restriction to bounded ranking functions. It is easy to construct by induction a formula $\mathit{reach}_B(\mathbf{C}, \mathbf{C}')$  that holds if there exists  a configuration $C'$ reachable from $C$ in at most $B$ steps. Just take 
\begin{eqnarray*}
\mathit{reach}_1(\mathbf{C}, \mathbf{C}') & := & \mathbf{C} = \mathbf{C}' \vee \bigvee_{t \in T} \mathit{step}_t(\mathbf{C}, \mathbf{C}') \\
\mathit{reach}_{k+1}(\mathbf{C}, \mathbf{C}') & := &  \exists \mathbf{C}'' \colon \mathit{reach}_{1}(\mathbf{C}, \mathbf{C}'') \wedge 
\mathit{reach}_{k}(\mathbf{C}'', \mathbf{C}')
\end{eqnarray*} 
\noindent Now we can express the weakly decreasing property as above, replacing $\mathit{reach}$ by $\mathit{reach}_B$.

As we mentioned before, the proof of existence of Presburger stage graphs is based on deep results on the theory of Petri nets, which can also be applied to population protocols. The main one is Leroux's theorem \cite{Leroux12}, stating the following. Let $X$ and $Y$ be Presburger sets of configurations of a Petri net, and let $\mathit{reach}(X)$ be the set of configurations reachable from $X$. The theorem states that if $\mathit{reach}(X) \cap Y = \emptyset$ holds, then there exists a Presburger inductive invariant that certifies this fact, i.e., there exists a Presburger set $S$ closed under the reachability relation such that $\mathit{reach}(X) \subseteq S$ and $S \cap Y = \emptyset$.  Observe that if $\mathit{reach}(X)$ were always itself a Presburger set, then we could just take $S = \mathit{reach}(X)$.  Intuitively, Leroux's theorem shows that, while $\mathit{reach}(X)$ is not always a Presburger set, it is always very close to it (in fact, Leroux's proof shows that $\mathit{reach}(X)$ always belongs to a class of sets called \emph{almost semilinear}).

\subsection{Population protocols decide exactly the Presburger predicates}

Angluin \etal\ proved in \cite{AAER07} that population protocols compute exactly the Presburger predicates. The difficult part is to show that population protocols \emph{only} compute Presburger predicates. Let us show that this is a simple corollary of the fact that correctness can always be certified by Presburger stage graphs.

Consider a protocol that decides a predicate, say $\varphi$. Let $I$ be the set of initial configurations of the protocol, and let $I_1$ and $I_0$ be a partition of $I$ into the set of initial configurations that satisfy and do not satisfy $\varphi$, respectively. Our theorem shows that there exists a Presburger stage graph with initial stage $\Stage$ such that $I_1 \subseteq \Stage$. This stage graph proves that a run starting at any configuration of $\Stage$ eventually reaches stable consensus 1 w.p.1. Since the protocol decides $\varphi$, no configuration of $I_0$ belongs to $\Stage$, i.e., we have $\Stage \cap I_0 = \emptyset$. Together with  $I_1 \subseteq \Stage$, we have $\Stage \cap I = I_1$. But $\Stage$ is Presburger, and so is $I$ (indeed, $I$ is just the set of configurations with $0$ agents in non-initial states). Since Presburger sets are closed under intersection, $I_1$ is also Presburger.

\subsection{Automatic computation of stage graphs}

We have developed a practical approach to the computation of stage graphs, implemented in the tool \textsc{Peregrine} 2.0 \cite{EsparzaHJM20}. The design of the tool is guided by the theoretical results on stage graphs, and by the notion of \emph{dead transitions}. A transition $t$ is \emph{dead at a configuration} $C$ if no run starting at $C$ executes $t$, and $t$ is \emph{dead at a stage} $\Stage$ if $t$ is dead at every $C \in S$. Population protocols designed by humans usually run in phases. Initially, all transitions are alive, and the end of each phase is marked by the ``death'' of one or more transitions, i.e., by reaching a configuration at which these transitions become dead. Runs of the protocol keep ``killing transitions'' until they reach a consensus configuration whose consensus cannot be broken by any of the transitions still alive.  This consensus is then stable. When applied to the majority protocol, \textsc{Peregrine} 2.0 computes automatically two stage graphs very similar to those of Figure \ref{fig:stagegraph} in a couple of seconds. 

Like \textsc{Peregrine} 1.0,  \textsc{Peregrine} 2.0 is built on top of the Z3 constraint solver. More precisely, it uses Z3 to check satisfiability of formulas of the existential fragment of Presburger arithmetic. The existential fragment is as expressive as full Presburger arithmetic, but can be handled much more efficiently.

Given a protocol and a Presburger predicate, \textsc{Peregrine} 2.0 computes two stage graphs, proving that runs starting at every initial configuration of the protocol satisfying (resp. violating)  the predicate eventually reach stable consensus $1$ (resp. stable consensus $0$) w.p.1.
Let $I_1$ be the set of initial configurations satisfying the predicate, the other case being similar. \textsc{Peregrine} 2.0
maintains a worklist of Presburger stages, finitely represented by Presburger formulas. Initially, the worklist contains only one stage, namely an inductive Presburger overapproximation $\PReach(I_1)$ (for ``potentially reachable'') of the configurations reachable from $I_1$. The procedure computing  $\PReach(I_1)$ is the result of many years of research on tractable ``relaxations'' of the reachability relation of Petri nets \cite{EsparzaM00,ELMMN14,FracaH15,BFHH17,Blondin20}. 

In its main loop, \textsc{Peregrine} 2.0 repeatedly picks a Presburger stage $\Stage$ from the worklist, and processes it. First, the tool checks whether $\Stage$ is terminal, \ie, if all its configurations are a $1$-consensus. (Since $\Stage$ is inductive by construction, an affirmative answer implies that all configurations of $\Stage$ are stable $1$-consensuses.) Checking that every configuration of $\Stage$ is a 1-consensus reduces to checking unsatisfiability of a simple formula. If $\Stage$ is not terminal, the tool attempts to construct one or more successor stages with strictly more dead transitions than $\Stage$.  For this, the tool computes a set of \emph{eventually dead transitions}: transitions that are alive at one or more configurations of $\Stage$, but will become dead w.p.1 in any run starting at those configurations. Again, the procedure to compute $U$ makes heavy use of results of Petri net theory \cite{JP19}, but also of the theory of well-quasi-orders \cite{ACJT96,FS01}. 

If \textsc{Peregrine} 2.0 finds a nonempty set of eventually dead transitions, then it constructs a successor stage of $\Stage$ by overapproximating the configurations reachable from $\Stage$, underapproximating the configurations at which the transitions of $U$ are dead, and intersecting the results. If \textsc{Peregrine} 2.0 fails to find eventually dead transitions, it heuristically splits $\Stage$ into different stages and adds them to the worklist to be processed.  Indeed, it could be the case that no transition becomes eventually dead from \emph{every} configuration of $\Stage$, but this no longer holds after a split; for example, imagine that transition $t_1$ eventually becomes dead from every configuration of $\Stage_1 \subset \Stage$, and another transition $t_2$ becomes eventually dead from every configuration of $\Stage_2 = \Stage \setminus \Stage_1$. In this case, after splitting $\Stage$ into $\Stage_1$ and $\Stage_2$ the tool can find nonempty sets of eventually dead transitions for both $\Stage_1$ and $\Stage_2$.

\textsc{Peregrine} 2.0 has successfully proved correct a large variety of protocols, including majority and approximate majority protocols (\Cref{ex:majority}, \cite[Ex.~3]{BEK18}, \cite{AGV15}, \cite{Cardelli2012}, \cite[\emph{coin game}]{LLMR17}, \cite{Mor58}), various \emph{flock-of-birds} protocol families (\cite{CMS10}, \cite[Sect.~3]{BEJ18}, \cite[\emph{threshold-n}]{CDFS11}) for the family of predicates $x \geq k$ for some constant $k \geq 0$; or protocols for threshold and remainder predicates of~\cite{AADFP04,BEJ18}. For all these examples \textsc{Peregrine} 2.0 computes stage graphs with a few stages. Currently, the main limitation of the tool is the size of the systems of linear constraints involved, which limits the tool to protocols with up to some dozens of states and some thousands of transitions.


\section{Succinct predicates and state complexity}
\label{sec:succinct}

After writing our first papers on the verification of population protocols, we observed that the theory of Petri nets was also relevant for a problem that,
perhaps surprisingly, had not been studied yet: the \emph{state complexity} of predicates decidable by population protocols. Informally, the state complexity of a predicate is the minimal number of states of the protocols that decide it, and, given a number $\eta$, one defines the function $\STC(\eta)$ as the maximum state state complexity of all predicates of size at most $\eta$. But what is the size of a predicate? This requires us to fix a representation. Since population protocols compute exactly the predicates expressible in Presburger arithmetic \cite{AAER07},  we must choose a representation of the Presburger sets. There are three natural representations: formulas of Presburger arithmetic, quantifier-free formulas of Presburger arithmetic, and semilinear sets \cite{Haase18}. Semilinear sets
are difficult to parse by humans, and no paper on population protocols uses them to describe predicates. Full Presburger
arithmetic is very succinct, but the complexity of its satisfiability problem lies between 2-NEXP and 2-EXPSPACE \cite{Haase18}, and so it can lead to results in which a predicate requires very few states, but only because of a representation that is very difficult to compute.
This leaves quantifier-free Presburger arithmetic.  This representation also has two advantages
of its own. First, standard predicates for which numerous protocols have been given in the literature  (like majority, threshold, or remainder predicates) are naturally expressed without quantifiers. Second, the procedures given so far to construct population protocols for any given  Presburger predicate explicitly use the fact that Presburger arithmetic has a quantifier elimination procedure, i.e., they first construct protocols for all threshold and remainder predicates, and then show that the predicates computed by protocols are closed under negation and conjunction. 

\subsection{State complexity: upper bounds}  The first synthesis procedure for the construction of a protocol deciding a given Presburger predicate was presented in \cite{AAE06}. The procedure is simple and elegant, but it yields large protocols. Given a quantifier-free Presburger formula $\varphi$, \ie, a boolean combination of atomic formulas, the number of states of the synthesized protocol grows exponentially in both the number of bits of the largest coefficient of $\varphi$ in absolute value, and the number of atomic formulas. In terms of $|\varphi|$ (defined as the number of bits needed to write $\varphi$, with coefficients written in binary) they have $\Omega(2^{\poly(|\varphi|)})$ states. This raises the question whether protocols with  $O(\poly(|\varphi|))$ states, which we call \emph{succinct}, exist.  We gave an affirmative answer in \cite{BEGHJ20}, completing first partial results obtained in \cite{BEJ18}. We describe how to avoid both exponential dependencies.

\paragraph{Handling large coefficients.} In order to prevent having the exponential dependence on the coefficients, we design protocols for threshold and remainder predicates that, loosely speaking, represent numbers in binary. A very easy case is described in Example \ref{ex:silent}: the first predicate for $x - y \geq 2^k$ has $\Theta(2^k)$ states, because agents can hold any number of pebbles between $0$ and $2^k$, but the second has only $\Theta(k)$ states, because the number of pebbles is always a power of $2$. The construction of \cite{BEGHJ20} proceeds in two steps: first we construct a succinct protocol in which the agents are assisted by \emph{helpers}, additional agents that are not part of the input, and initially occupy a distinguished state, say $H$; then we give an equivalent protocol without helpers. Helpers are similar to leaders, but with the property (guaranteed by the design of the protocol) that if the protocol works correctly with a certain number of helpers, then it also works correctly for any larger number. This property is crucial when dealing with boolean combinations.

Consider a predicate like $13 x - 9y \geq 5$. Protocols for this predicate have two initial states, for $x$ and $y$. Since $13 = 2^3 + 2^2 + 2^0$, the protocol has a transition that moves an agent in the initial state for $x$ and two helpers into states $2^3$, $2^2$, and $2^0$, respectively. (Transitions  that move multiple agents to new states in one single step can be easily simulated by the usual protocol transitions with a very low cost in terms of states.) Similarly, another transition for each agent in the initial state for $y$, the protocol moves this agent and one helper into states $-2^3$ and $-2^1$. Pairs of agents in states $2^l$ and $-2^l$ can ``cancel'', meaning that they both move to state $H$, and so become helpers ready to continue assisting. 

In \cite{BEGHJ20} we show that this idea can be applied to every threshold or remainder predicate, resulting in a succinct protocol with a \emph{fixed} number of helpers, cubic in the size of the predicate, but independent of the size of the input. But how do we go from this protocol to another one without helpers?  For large inputs in which the number of agents exceeds this number of helpers, we can let each agent  take two jobs: act as a regular agent \emph{and} a helper. Let us show how to do for $h$ helpers. In a first phase, the protocol assigns to each agent a number between 1 and $h$, ensuring that each number is assigned to at least one agent (this is the point at which we need a sufficiently large input with at least $h$ agents). More precisely, at the end of this phase each agent is in a state of the
form $(x, i)$, meaning that the agent initially represented one
unit of input for variable $x$, and that it has been assigned
number $i$.  For this, initially every agent is placed in
    state $(x, 1)$. Transitions of the form $(x, i),
      (x, i) \mapsto (x, i+1), (x, i)$ for every $1 \leq
    i \leq h-1$ guarantee that all but one agent is promoted to $(x, 2)$, all but one to $(x, 3)$, etc. In other
    words, at each step one agent is ``left behind'', and so the protocol has at least $h$ helpers. 

However, the protocol must be correct for all inputs, not only for those with at least $h$ agents. In \cite{BEGHJ20} this is solved by designing a second family of protocols for small inputs, which works in a completely different way. It is then easy to combine the protocols for large and small inputs into a protocol for all inputs. 

\paragraph{Handling large boolean combinations of atomic formulas.} The second problem of the synthesis procedure of \cite{AAE06} is the exponential dependence of the number of states on the number of atomic formulas. The dependence comes from the fact that, given protocols $\PP_1, \ldots, \PP_k$ with $n_1, \ldots, n_k$ states deciding formulas $\varphi_1, \ldots, \varphi_k$, respectively,  the synthesis procedure yields a protocol $\PP$ for deciding $\varphi_1 \wedge \cdots \wedge \varphi_k$ with $n_1 \cdot n_2 \cdot \ldots \cdot n_k$ states (and similarly for $\varphi_1 \vee \cdots \vee \varphi_k$). Intuitively, in  $\PP$ each agent carries out $k$ jobs: act as an agent of $\PP_1$,  of $\PP_2$, \ldots, and of $\PP_k$. The state of an agent is a $k$-tuple of states of the $\PP_1, \ldots, \PP_k$, and when two agents meet, they compare their states in each protocol $\PP_i$, and apply the corresponding transition. In other words: the new protocol executes all of $\PP_1, \ldots, \PP_k$ synchronously.

We need a new succinct construction for a boolean
combination of atomic predicates with $O(n_1 + \cdots + n_k)$ instead of $\Omega(n_1 \cdot \ldots \cdot n_k)$ states.  A naive first idea is to let $\PP$ execute $\PP_1, \ldots, \PP_k$ asynchronously in parallel, instead of synchronously, and combine the results.  However, this does not work.  Assume $\varphi_1, \ldots, \varphi_k$ have arity $m$.
In order to compute  $(\varphi_1 \wedge \cdots \wedge \varphi_k)(\vec{x})$, where $\vec{x}=(x_1, \ldots, x_m)$, the protocol $\PP$ would have to dispatch $\vec{x}$ agents to (the input states of) each $\PP_i$, giving a total of $k \cdot \vec{x}$ agents, but the protocol only has the agents of the input, i.e. $\vec{x}$ agents. But couldn't we use $(k-1) \cdot \vec{x}$ helpers, and then obtain an equivalent protocol without helpers? No, this does not work either, because in this case the number of helpers depends on the size of the input, and the technique we used above only allows us to simulate a fixed number of helpers. The solution is to use a more sophisticated construction for parallel asynchronous computation. Again, we need to consider separately the cases of large and small inputs, but the former is more interesting, and so we only describe protocols for large inputs. 

\newcommand{\firstcopy}{y}
\newcommand{\secondcopy}{z}

Given an arbitrary threshold or modulo predicate $\psi(\vec{x})$ of arity $m$, it is easy to construct a predicate $\widetilde{\psi}(\vec{\firstcopy},\vec{\secondcopy})$
of arity $2m$ satisfying 
$$\widetilde{\psi}(\vec{\firstcopy},\vec{\secondcopy}) = \psi(k\vec{\firstcopy}+\vec{\secondcopy})$$
For instance, if $\psi(x_1,x_2)= (3 x_1 - 2x_2 > 6)$
and $k=4$, then we can choose
$\widetilde{\psi}(\firstcopy_1,\firstcopy_2,\secondcopy_1,\secondcopy_2) =  (12 \firstcopy_1 + 3 \secondcopy_1 -
8 \firstcopy_2 - 2 \secondcopy_2 > 6)$. 

Intuitively, the idea is to let $\PP$ compute  $\widetilde{\varphi}_1(\vec{y},\vec{z}), \ldots, \widetilde{\varphi}_k(\vec{y},\vec{z})$ instead of $\varphi_1(\vec{x}), \ldots, \varphi_k(\vec{x})$ for some $\vec{y}$ and $\vec{z}$ satisfying $\vec{x}=k\vec{y}+\vec{z}$. Then $\PP$ only needs to dispatch a total of  
$$k \left( \sum_{i=1}^m y_i + z_i \right) = k \left( \sum_{i=1}^m y_i + (x_i - k y_i) \right) \leq \sum_{i=1}^m x_i + m \cdot (k-1)^2$$
\noindent agents to compute all of $\widetilde{\varphi}_1, \ldots, \widetilde{\varphi}_k$. So $\PP$ only needs $m \cdot (k-1)^2$ helpers, a fixed number independent of the number of agents.

Let us now describe how  $\PP$ computes $\widetilde{\varphi}_i(\vec{y},\vec{z})$ for some 
$\vec{y}$ and $\vec{z}$ satisfying $\vec{x}=k\vec{y}+\vec{z}$.
Let $\widetilde{\PP}_1, \ldots, \widetilde{\PP}_k$ be protocols computing $\widetilde{\varphi}_1, \ldots, \widetilde{\varphi}_k$, 
let $\q{x_1}, \ldots, \q{x_m}$ be the input states of $\PP$, 
and let
$\q{\firstcopy^j_1}, \ldots, \q{\firstcopy^j_m}$ and $\q{\secondcopy^j_1}, \ldots, \q{\secondcopy^j_m}$ be the input states of $\widetilde{\PP}_j$
for every $1 \leq j \leq k$. 
Protocol $\PP$ 
repeatedly chooses an index $1 \leq i \leq m$, and executes one of these two actions, which can be implemented with some effort using only binary interactions: take $k$ agents from $\q{x_i}$, and dispatch them to $\q{\firstcopy^1_i}, \ldots, \q{\firstcopy^k_i}$ (one agent to each state); or take one agent from $\q{x_i}$ and $(k-1)$ helpers, and dispatch them to $\q{\secondcopy^1_i}, \ldots, \q{\secondcopy^k_i}$. 
If all agents of $\q{x_i}$ are dispatched for every $1 \leq i \leq m$, then we say that the {\em dispatch is correct}. Observe that
a correct dispatch satisfies $\vec{x}=k\vec{y}+\vec{z}$. 

The problem is that the dispatch may or may not be correct.
Assume, \eg, that $k=5$ and $m=1$. Consider the input $x_1=17$, and assume that $\PP$ has $m \cdot (k-1)^2=16$ helpers. $\PP$ may correctly dispatch
$\firstcopy_1 =  3$ agents to each of $\q{\firstcopy^1_1}, \ldots, \q{\firstcopy^1_5}$
and $\secondcopy_1 = 2$ to each of $\q{\secondcopy^1_1}, \ldots, \q{\secondcopy^1_5}$; this gives a total of $(3+2)\cdot 5 = 25$ agents, consisting of the $17$ agents for the input plus $8$
helpers. However, it may also wrongly dispatch $2$ agents to each $\q{\firstcopy^1_i}$ and
$4$ agents to each of $\q{\secondcopy^1_i}$, with a total of $(2+4) \cdot 5 = 30$ agents, consisting of $14$ input agents plus $16$ helpers. In the second case, each $\PP_j$
wrongly computes $\widetilde{\varphi}_j(2, 4) = \varphi_j(2\cdot 5 + 4) = \varphi_j(14)$, instead of the correct value $\varphi_j(17)$.

To solve this problem we ensure that $\PP$ can always recall agents already dispatched 
to $\widetilde{\PP}_1, \ldots, \widetilde{\PP}_k$ as long as the dispatch is not yet correct. 
This allows $\PP$ to ``try out'' dispatches until it dispatches correctly, which happens eventually w.p.1. 
For this we design $\PP$ so that the atomic protocols $\widetilde{\PP}_1, \ldots, \widetilde{\PP}_k$ can work with  agents that arrive to the initial states over time ({\em dynamic initialization}), and can always return to their initial states and go back to $\PP$, unless the  dispatch is correct ({\em reversibility}).
To ensure that $\PP$ stops recalling agents after a correct dispatch, we modify the dispatch transitions so that they  become disabled when $\q{x_1}, \ldots, \q{x_m}$ are not populated.

\subsection{State complexity: lower bounds}  In \cite{BEJ18,CzernerE21} we have also studied the problem of obtaining lower bounds for $\STC(\eta)$.  This question turns out be surprisingly hard, and so we have focused on obtain lower bounds for the state complexity of a particularly simple family of predicates, namely those of of the form $x\geq k$.  This amounts to finding a lower bound for the number $n$ of states needed to decide $x \geq k$, or, equivalently, an upper bound for the largest number $k$ such that $x \geq k$ can be decided by a protocol with $n$ states. We prefer the latter formulation due to its analogy with the busy beaver function. Recall  that the busy beaver function assigns to a number $n$ the largest $\eta$ such that a Turing machine with at most $n$ states, started on a blank tape, writes $\eta$ consecutive 1s on the tape and halts. Analogously, the busy beaver function for population protocols assigns to $n$ the largest $\eta$ such that a population protocol with at most $n$ states  decides the predicate $x \geq \eta$. Intuitively, $\eta$ is the largest number ``recognizable'' by protocols with at most $n$ states. 

We have obtained results for protocols with and without leaders. It is known that the time complexity of predicates is different for population protocols with and without leaders: While the first can decide any Presburger predicate in poly-logarithmic parallel time \cite{AngluinAE08a},  the latter need linear parallel time for majority \cite{AlistarhAG18}. Is the same true for state complexity? The question is still open, but we have made some progress.

Let  $\BB, \BBL \colon \N \rightarrow \N$ be the busy beaver functions for leaderless protocols and for protocols with leaders, respectively. A protocol similar to the second one of Example \ref{ex:silent}, only simpler, decides $x \geq 2^n$ with $O(n)$ states, showing that $\BB(n) \in \Omega(2^n)$. In \cite{BEJ18} we prove $\BBL(n) \in \Omega(2^{2^n})$. 
This result is quite surprising: for certain numbers $k$, there are population protocols that decide $x \geq k$ even though an agent does not have enough memory to index even one bit of $k$. The proof follows from a theorem by Mayr and Meyer on presentations of commutative semigroups \cite{mayr1982complexity}, which can be reformulated in protocol terms as follows: for every $n \geq 1$, there exists a protocol with $O(n)$ states and three distinguished states $\mathit{start}, \mathit{end}, \mathit{counter}$  such that from an initial configuration that puts one agent in $\mathit{start}$ and $k$ agents in $\mathit{counter}$, respectively, it is possible to reach a configuration putting at least one agent in state $\mathit{end}$ if and only if $k \geq 2^{2^n}$.  By adding transitions that allow an agent in state $\mathit{end}$ to attract all other agents to $\mathit{end}$, it is easy to obtain a protocol deciding $x \geq 2^{2^n}$.

Can we also obtain upper bounds on $\BB(n)$ and $\BBL(n)$, and so lower bounds on the state complexity? After some years investigating this question without progress, we have recently made a breakthrough \cite{CzernerE21}. Our first result is that $\BBL(n)$ is bounded by a variant of the Ackermann function. The result is proved by means of a pumping technique. We first show that if a protocol with $n$ states answers $0$ for two inputs $a < b$ satisfying certain conditions, then it also answers $0$ for every input $a + \lambda (b -a)$, and so the protocol cannot compute any predicate of the form $x \geq \eta$. Then we find a function $F(n)$ such that if a protocol with at most $n$ states rejects all inputs with at most $F(n)$ agents, then there are inputs $a < b < F(n)$ satisfying the conditions. The function is obtained using results from the theory of controlled sequences, an area of mathematics related to well-quasi-orders \cite{AbriolaFS15,Balasubramanian20,FigueiraFSS11}.

An Ackermannian upper bound may seem extremely weak. However, it follows from recent results in the theory of Petri nets that functions similar to $\BBL(n)$ have an Ackermannian \emph{lower} bound. To give an example, say that a protocol \emph{weakly decides} the predicate $x \geq k$ if the following holds: for every initial configuration with at least $k$ agents there exists a run leading to a configuration with consensus $1$; for every initial configuration with less than $k$ agents, no such configuration is reachable. Then the largest $k$ such that $x \geq k$ is weakly computable with $n$ states is an Ackermannian function of $n$. 

The main result of \cite{CzernerE21} is a triple exponential bound on $\BB(n)$. That is, leaderless protocols with at most $n$ states can recognize numbers at most triple exponential in $n$. The proof technique is again a pumping lemma. The key property of leaderless protocols that we use to obtain an elementary bound is that, loosely speaking, the set of initial configurations of a leaderless protocol is closed under addition. To understand this, observe that  initial configurations of a leaderless protocol deciding $x\geq k$ put $k$ agents in the initial state and $0$ agents in all others. Therefore, the sum of two initial configurations with $k_1$ and $k_2$ agents is the initial configuration with $k_1 + k_2$ agents. This does no longer hold for protocols with a leader, whose initial configurations also put one agent in the initial state of the leader; in this case, the sum of two initial configurations with a leader is a configuration with \emph{two} leaders.

In an unpublished result obtained together with J\'er\^ome Leroux, we have improved the bound for leaderless protocols to double exponential; we conjecture that the optimal upper bound is single exponential, matching the lower bound. But currently we do not even have a line of attack to obtain an elementary bound for protocols with a leader.

\section{Conclusions and future work}
We have surveyed our recent work on the verification of population protocols, and on their state complexity. This work has produced \textsc{Peregrine}, the first automatic tool able to verify correctness of protocols for all inputs. In the verification area, there are many open directions for future work: 
\begin{itemize}
\item Protocols are often designed parametrically, for example, one gives a construction that yields a protocol deciding $a x - by \geq c$ for arbitrary coefficients $a, b, c$. Our methods cannot yet prove that the construction is correct for every $a, b, c$. It is easy to show that verifying infinite families of protocols is an undecidable problem, even for very restricted cases, but it should be possible to design procedures that perform well in practice. 
\item As mentioned in the introduction, in the last years families of protocols that decide one single predicate, but where the number of states increases with the number of agents, have been intensely investigated, see e.g. \cite{AlistarhG18,ElsasserR18}. Again, we do not have any verification technique for them. 
\item The work initiated in \cite{BEK18} on the automatic verification of the expected runtime of a protocol is still in its infancy.
\end{itemize}

Our work on the state complexity problem is tightly linked to difficult problems of the theory of Petri nets. The obvious future direction is closing the current gaps between the upper and lower bounds for the busy beaver functions in the leaderless case, and the case with leaders. We consider this a fundamental problem in the theory of population protocols. Intuitively, it measures quantitatively the relation between the microscopic scale of agents and the macroscopic scale of the predicates they decide. 

\bibliographystyle{splncs04}
\bibliography{rp21refs}

\end{document}